# Role of diffusion-induced grain boundary migration on molten salt corrosion of a Ni-30Cr alloy

Konnor Walter[1], Jagadeesh Sure[2], Adrien Couet[2], Emmanuelle A. Marquis[1,*]

[1 -] Department of Materials Science and Engineering, University of Michigan, Ann Arbor

[2 -] Department of Nuclear Engineering, University of Wisconsin, Madison

*Corresponding author (emarq@umich.edu)

**Abstract**

The response of Ni-Cr alloys to exposure to molten chloride and fluoride salts is typically characterized by Cr dealloying with the formation of a Cr-depleted bi-continuous porous subsurface layer. The exact mechanism behind the loss of Cr over distances unattainable by lattice diffusion alone is still debated. To address this question, two different surface finishes, namely electropolished and sanded, of a Ni-30Cr alloy were exposed to LiCl–KCl-2wt% $EuCl_3$ eutectic salt at 500 °C for 96 hours. In the absence of fast diffusion pathways, dissolution occurred layer by layer and was kinetically controlled by Ni dissolution, as observed over the grain interiors of the electropolished sample. Grain boundaries were subject to diffusion-induced grain boundary migration (DIGM), leading to the formation of pure Ni islands above grain boundaries. This overall behavior contrasted with the sanded surface response that was characterized by several micrometer deep interconnected porosity and complete Cr depletion. DIGM of the dense grain boundaries created by recrystallization of the sanded surface was responsible for the observed sub-surface microstructure. This work unequivocally establishes DIGM as a key mechanism in alloy molten salt corrosion, and microstructure as a decisive contributor to an alloy's corrosion response.

# Introduction

Molten salt reactors (MSRs) are central to the development of next generation (Gen IV) nuclear power plants [1–4]. Molten salts have been proposed as fuels and secondary coolants for MSRs and electrolytes for the pyrochemical reprocessing of spent fuels [5]. Beyond nuclear applications, molten salt technology is also being explored for concentrated solar power plants [6,7]. Among various salt systems, molten chloride salts have gained significant interest due to their excellent thermophysical properties, making them suitable for energy applications. However, a major challenge is the corrosion of structural materials [8], leading to the deterioration of properties and structural integrity when these materials are exposed to molten salts under extreme conditions [9]. Whether and how a material corrodes can be strongly influenced by the chloro-acidity of the salt mixtures and the presence of impurities [10][11].

Fe-based and Ni-based alloys are considered candidate materials for structural and cladding applications [12]. These alloys undergo degradation in molten chloride salts, resulting in complex interfacial reactions at the alloy/salt interface and producing various forms of corrosion attack. Studies on the corrosion behavior of these alloys in molten salts clearly indicate that the extent of corrosion strongly depends on the alloy composition. In commercial Cr-containing Fe- and Ni-based alloys, chromium plays a critical role and can undergo selective dissolution, often leading to the formation of porous or void regions near the alloy surface. To gain a deeper understanding of Cr dissolution and the associated microstructural changes at localized surface and subsurface regions, many studies have focused on binary Ni-Cr model alloys [12–23].

One of the primary corrosion mechanisms for Ni-Cr model alloys in molten salts has primarily been reported to take the form of dealloying and subsurface pore formations [8,13–15]. Dealloying occurs when one of the elements (referred to as solute herein) selectively dissolves into the liquid medium, such as an electrolyte, molten salt, or liquid metal [16,17,24]. In the case of Ni-Cr based alloys in oxygen-deficient halide melts, Cr can selectively dissolve leaving behind a Ni-rich surface due to the greater electrochemical stability of Ni [13,18]. At intermediate temperatures where lattice diffusion is limited, it is thought that the Ni atoms left at the alloy surface reorganize through rapid surface diffusion. This combination of selective dissolution and surface diffusion leads to the formation of so-called 'bi-continuous structures' of ligaments and pores [16,17,24]. The characteristic pore size and morphology are then governed by the competition between dissolution kinetics and surface diffusion rates [16,17,24]. In addition to bi-continuous structures, other corrosion morphologies have also been reported, such as the absence of selective dissolution [19], severe grain boundary attack, and corrosion taking place through cracks [20,21]. The exact response depends on the salt and alloy composition, salt's redox potential [19,25], temperature, and alloy composition impurity content [22].

During the formation of the commonly observed bi-continuous pore/ligament structures, solute depletion below the surface has been observed over distances far exceeding those predicted by lattice diffusion alone, e.g., [15,18,23]. This discrepancy indicates that fast diffusion pathways, particularly grain boundaries, are required to transport solutes to the surface [18,23]. Indeed, increasing the density of such paths has been linked to accelerated corrosion [25–27]. Prior studies also proposed that solutes diffuse through the lattice before reaching grain boundaries along which they can then rapidly diffuse to the surface [15,18]. However, this framework does not explain how solutes are able to reach grain boundaries in systems with large grains. Indeed, solute depletion has been noted over distances away from grain boundaries that are many orders of magnitude larger than the expected diffusion distance.

It was recently suggested that diffusion-induced grain boundary migration (DIGM) may play a role in molten salt corrosion of Ni-Cr alloys [28]. DIGM describes the process by which solute diffusion in the grain boundary plane couples with migration of the grain boundary [29,30]. In the case of selective corrosion, it is characterized by the creation of dealloyed regions in the wake of the moving grain boundaries. Experimental evidence for DIGM in the context of alloy corrosion in molten salt environments was reported by Teng et al. [31], Yang et al. [32], and Mills et al. [33] who found the formation of pores along migrated grain boundaries in Ni-20Cr corroded at 600 °C and 650 °C in FLiNaK salt. Cr depletion was noted behind the migrated portions of the grain boundaries.

Given the role of fast diffusion pathways, it is relevant to revisit the impact of surface deformation, introduced during surface preparation or cold working prior to corrosion exposure, on the corrosion behavior of alloys in molten salts. Increasing the density of starting defects (dislocations and grain boundaries) was generally found to enhance the corrosion susceptibility of steels and Ni alloys, which most authors rationalized via the high density of fast diffusion paths able to bring Cr to the surface [25,34]. Chan et al. [35] reported that increasing the degree of cold-rolling induced deformation in Ni-20Cr alloys increased the development of a bicontinuous porous structure when tested in FLiNaK molten salt at an applied potential in the active potential range. This bicontinuous structure, present both within grain interiors and along grain boundaries, was attributed to the high thermodynamic driving force for Cr dissolution and defect-mediated outward solid-state diffusion of Cr. It is, however, worth noting that at the temperatures used for the molten salt exposures, some alloys, depending on their degree of cold working, may experience rapid recovery, recrystallization, and grain growth. It was for example the case for a Ni-30Cr alloy whose surface was sanded, i.e., deformed over a few microns of depth, and exhibited recrystallization and grain growth within 10 minutes at 600 °C, enabling significant Cr diffusion to the alloy surface via extensive DIGM driven during air oxidation [36]. Therefore, understanding the role of surface preparation, starting alloy grain size, and the impact on cold rolling requires understanding the synergy between the evolving alloy microstructure and the corrosion processes.

In this context, we revisited the role of surface or sub-surface microstructural defects in the corrosion behavior of a model Ni-30Cr alloy to clarify how grain boundaries contribute to the selective dissolution of Cr in molten salt. To address this, we conducted molten salt droplet corrosion studies on electropolished and sanded surfaces of Ni-30Cr alloys in LiCl–KCl–2 wt.% $EuCl_3$ salt at 500 °C. The dissolution behavior of grains and grain boundaries, as well as the role of DIGM in Cr transport, was further investigated using electron microscopy characterizaton. We show that the presence of initial plastic deformation alters the corrosion process, highlighting the potential impact of the final surface finish and manufacturing history of a component.

**Methods**

A Ni-30 at.% Cr alloy was prepared by arc melting high purity Ni and Cr. The ingots were flipped and remelted five times to ensure chemical homogeneity, then homogenized at 1100 °C in Ar for 24 hours. The alloy grain size averaged 240 µm. Corrosion samples were sectioned using a low-speed saw and cleaned with 600-grit sandpaper. A subset of samples received additional polishing with 0.05 µm diamond paste followed by electropolishing in a 20% perchloric acid/methanol solution at -50 °C at 30 V for 30 s. The remaining samples were polished with a 240-grit sandpaper. These two distinct surface preparations are hereafter referred to as 'sanded' and 'electropolished' (EP) surfaces. These two surface conditions influence the extent of sub-surface deformation during the preparation process, as previously discussed, e.g. [36].

The Ni-30Cr samples (11mm x 11mm x 1mm squares) were cleaned, vacuum dried, and transferred to an Ar filled glove box for corrosion studies. Anhydrous LiCl and KCl (> 99% purity, Sigma Aldrich) were used to prepare the eutectic 44 wt% LiCl-56 wt% KCl mixture and 2 wt% $EuCl_3$ (99.99% purity, Sigma Aldrich) was added. The salts were weighed and mixed inside the glove box using a mortar and pestle, then die-pressed into pellets (~0.28 g each) using a tungsten carbide die and compacted with a Pike Pixie Manual Hydraulic Press under an applied load of 2.5 tons for 2 minutes into 8.5 mm diameter and 3 mm height cylindrical discs. All salt handling was performed inside an argon atmosphere glove box ($O_2$ < 1 ppm and $H_2O$ < 0.1 ppm).

Molten salt droplet corrosion experiments were performed on a heating plate housed within a box furnace. The detailed experimental procedure of molten salt droplet corrosion and setup details can be found in [37,38]. The setup was leveled before testing, and a pressed salt pellet was positioned at the center of the alloy surface. To prevent molten salt leakage during heating, a Pyrex cage was installed around the pellet. The furnace temperature was raised to 500 °C while monitoring the sample surface with a thermocouple inserted above the Ni plate. Upon melting, the pellet formed a stable droplet on the alloy surface. Corrosion was allowed to proceed for 96 hours. After testing, the furnace was switched off and the droplet solidified inside the glove box. The solidified salt was recovered and sealed for chemical analysis. The tested Ni-30Cr samples were removed from the glove box, rinsed three times in demineralized water, cleaned in acetone, vacuum dried, and stored for microstructural characterization. Two identical Ni-30Cr samples with the same surface finish were tested to ensure reproducibility.

The pre- and post-corrosion salt samples, recovered from solidified salt droplets on the Ni-30Cr sample surfaces, were sealed in glass vials inside a glove box and analyzed at the Wisconsin State Laboratory of Hygiene to determine corrosion product concentrations using Inductively Coupled Plasma Mass Spectrometry (ICP-MS). The salt samples were stored in a sealed containers until digestion preparation. Each salt sample was weighed and transferred into tared Teflon microwave digestion vials. Digestion of the LiCl–KCl–2 wt% $EuCl_3$ salt samples was performed using a mixture of concentrated nitric acid (16 M) and hydrochloric acid (12 M). This mixture was placed in sealed Teflon microwave digestion vials and heated to 195 °C with a two-step ramp, maintaining this temperature for 25 minutes. After digestion, the samples were diluted to 50 mL with ASTM Type I reagent water. Two to three samples were analyzed from each purified and corrosion-tested salts to ensure reproducible results. The impurity concentrations in the pre-corrosion of LiCl-KCl-2 wt % $EuCl_3$ salt are presented in **Table 1**.

**Table 1.** Impurity concentration in the pre-corrosion LiCl-KCl-2 wt % $EuCl_3$ salt mixture used for pressing salt pellets. All values are in w.ppm.

| Element | Fe | Cr | Ni | Mn | Al | Ca | Na | Mg | P | S |
|---|---|---|---|---|---|---|---|---|---|---|
| Measured concentration | 0.02 | 1.0 | 0.1 | 0.1 | 3.8 | 0.2 | 16.5 | 7.5 | 11.2 | 32.3 |

Microstructure characterization was performed using a Thermo Fisher G5 Hydra PFIB for SEM imaging, energy-dispersive X-ray spectroscopy (EDS) mapping, and scanning transmission electron microscopy (STEM) and TEM lamella preparation. STEM-EDS and nano-beam diffraction were performed on a Thermo Fisher Talos F200X G2 microscope. Image processing was carried out with *ImageJ* v153, *TEAMS* v4.6.2001, and *Velox* v3.12.1.5. *Velox* quantification was completed with a 2px average prefilter and 1.5 Gaussian post filter.

## Results

The solidified post-corrosion salt samples collected from Ni-30Cr surfaces were analyzed using ICP-MS, as shown in **Figure 1**. Based on the measured concentrations of dissolved metal ions in the corrosion-tested salt, a notable difference in Ni and Cr content was observed. On average, the dissolved Cr in the salt was about 20% higher for the sanded surface compared to the EP surface. In contrast, the concentrations of Ni in both samples were relatively similar. The Cr concentrations were higher than those of Ni, which is expected because the standard Gibbs free energy of formation for $NiCl_2$/Ni (-188.346 kJ/mol) is much higher than that for $CrCl_2$/Cr (-298.301 kJ/mol) at the corrosion testing temperature of 500 °C, based on the thermodynamic database from HSC Chemistry 6.0 [39]. The concentration ranges observed for Ni and Cr in the salts were comparable to those reported for corrosion products in similar droplet molten LiCl-KCl-2wt % $EuCl_3$ salt corrosion studies conducted on Cr-Fe-Mn-Ni alloys [37,38]. The salt did not reach saturation with corrosion products. Indeed, the solubilities of $Ni^{2+}$ and $Cr^{2+}/Cr^{3+}$ in molten LiCl–KCl salt [40,41] are significantly higher than the measured metallic cation concentrations reported in **Figure 1**. This finding further supports our experimental approach, indicating that 96 hours of corrosion testing is sufficient for screening Ni-30Cr alloy samples with two different surface treatments to understand corrosion behavior in molten salts. Samples 1 and 3 were used for further characterization.

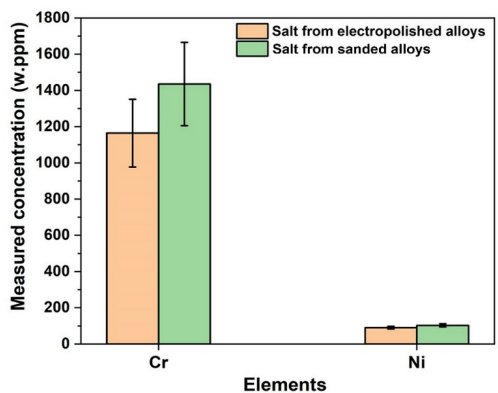

*Figure 1.* Average concentrations (measured by ICP-MS) of dissolved Ni and Cr metal ions in the post-corrosion salts.

SEM images of the surfaces before corrosion are provided in **Figure 2** for reference. As expected from the preparation methods, the electropolished surface was smooth, while the ground surface consisted of dense and parallel scratches. After corrosion, the samples showed very different morphologies depending on the initial surface condition. For the EP sample (**Figure 3a-d**), a continuous band at the surface, which SEM-EDS revealed to be pure Ni (**Figure 4a-c**), were formed above the alloy grain boundaries. The surface grain interiors were textured on a finer scale with facets that suggested a strong crystallographic dependence. For the sanded sample, a more uniform surface appearance was noted. Two types of contrast regions were also noted (**Figure 3e-f**), and further SEM-EDS analysis confirmed that the lighter regions were pure nickel while the darker regions contained some amount of Cr (**Figure 4**). Furthermore, the lighter regions had a high degree of surface porosity while the darker regions had less surface porosity. Grain boundaries corresponding to significantly smaller grains than the initial alloy grain size were also visible in the porous region as shown in **Figure 3g**.

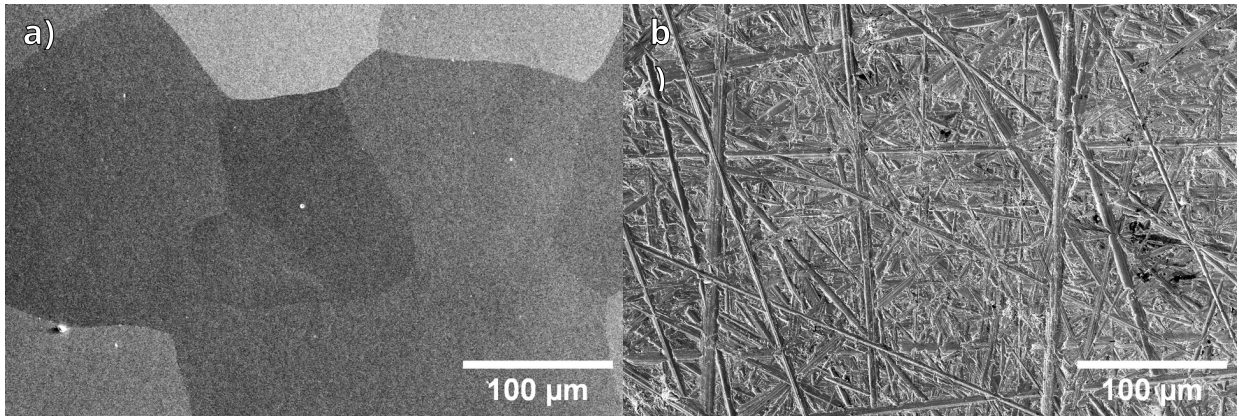

**Figure 2**: *SEM secondary electron micrographs of the surfaces of the (a) EP and (b) sanded specimens before corrosion.*

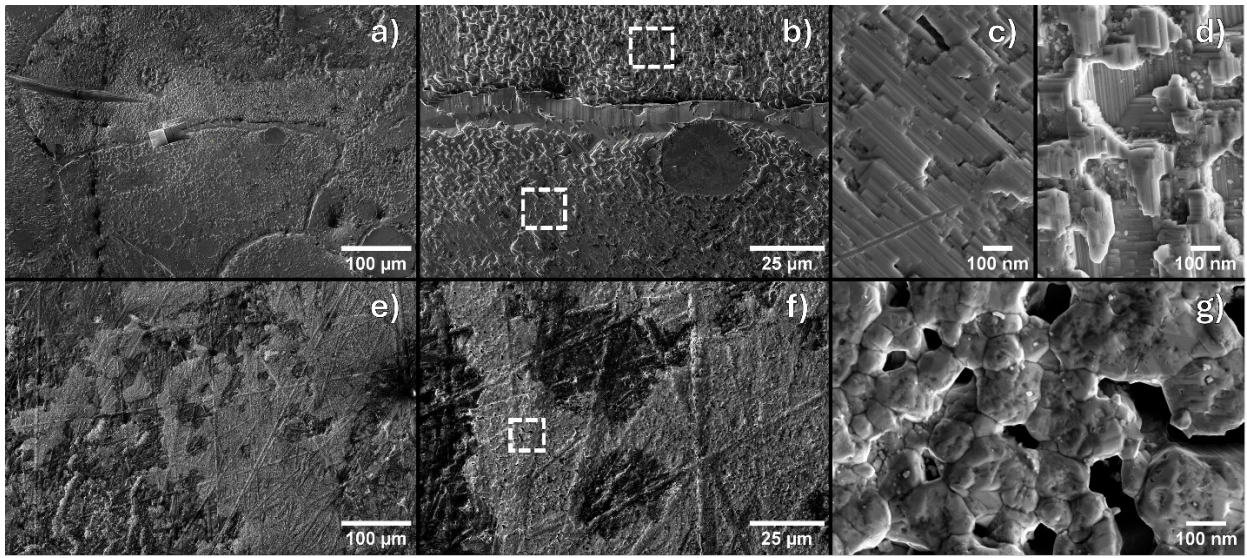

*Figure 3. Surface SEM images (secondary electrons) of the EP surface (top) and the sanded surface (bottom) after removal of the salt. (a, e) low magnification images of EP and sanded surfaces (b) Higher magnification image from the EP surface shown the different surface*

*morphology observed alloy grain boundaries. (c, d). Images taken at locations indicated in (b) and showing the different surface morphologies depending on the crystallographic orientation of the alloy surface. (f, g) higher magnification images of the surface morphologies found on the sanded surface.*

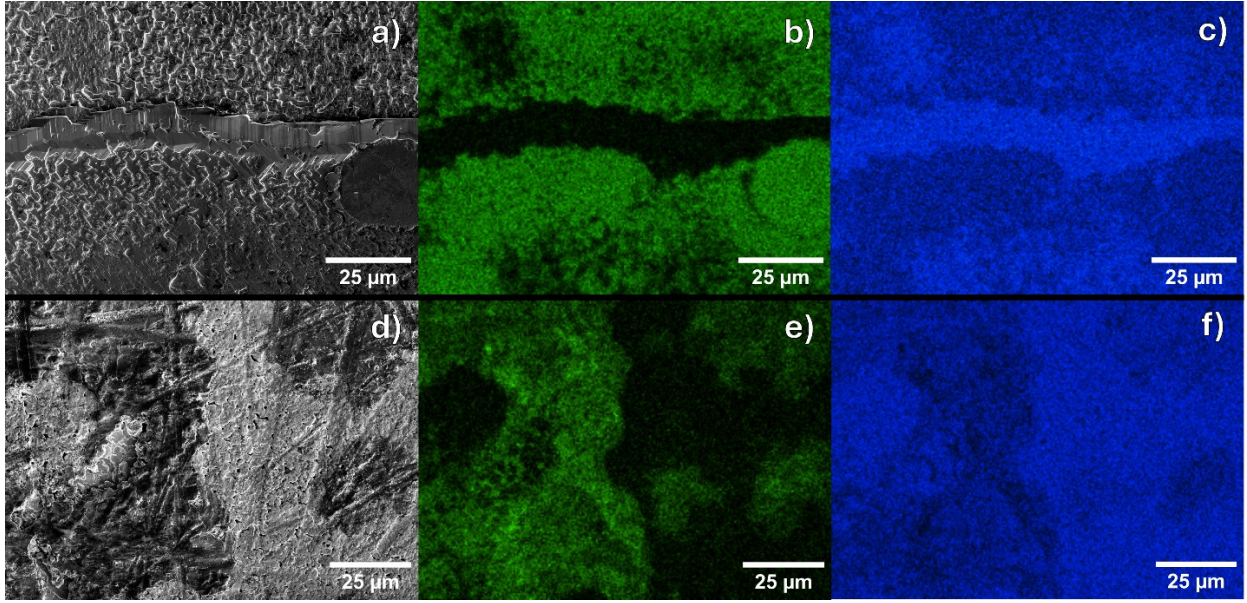

***Figure 4.*** *SEM images (secondary electron) and SEM-EDS maps taken from the EP (top) and sanded (bottom) surface after removal of the salt with corresponding EDS elemental maps for (b, e) Cr and (c, f) Ni.*

STEM observations of cross-sections were conducted to better understand the structure and chemistry of the surface and subsurface for each sample condition. For the EP sample (**Figure 5**), analysis was conducted at and around a high angle grain boundary (HAGB) exhibiting a protruding behavior at the corroded surface, such as the one shown in **Figure 4**. On either side of the grain boundary region, the grain interiors exhibited no preferential dissolution and the alloy chemistry was maintained all the way to the surface (**Figure 5b**). The protruding region above the alloy grain boundary was identified as nominally pure Ni. The HAGB displayed a shift in its position when transitioning from the alloy to the pure Ni region (left to right in **Figure 5**). A chemically abrupt interface was also noted between the pure Ni region and the region of nominal alloy composition, which electron diffraction analysis revealed to correspond to low angle grain boundaries (LAGBs). The left and right LAGBs had misorientations of 1° and 2° respectively, while the HAGBs had a misorientation of 39° in the pure Ni region and 38° in the alloy region.

Parts of the grain boundary in the alloy region had also moved (indicated by the dotted black arrow in **Figure 5a**). In the pure Ni region, the grain boundary was enriched in Cr while it was slightly depleted in Cr in the regions below, as shown via line profiles in **Figure 5c**.

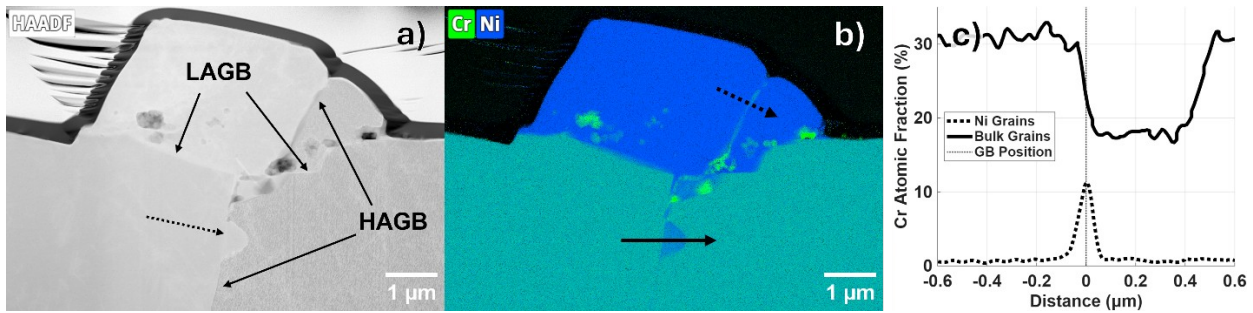

*Figure 5. (a) STEM image (HAADF) of a cross-section from the corrosion tested EP Ni-30Cr sample containing a grain boundary. (b) Corresponding EDS elemental map. (c) Cr composition profiles of the lines 1 and 2 indicated in (b), respectively.*

For the sanded sample, a very different microstructure was observed. First, the presence of Cr on the surface of the dark regions was attributed to a local thin Cr-rich oxide layer formed after corrosion during salt removal and sample handling. No microstructural differences were noted between the bright and dark regions; the observations were, therefore, considered representative of the entire surface regardless of their location. The cross-section analysis collected from the darker region in **Figure 3f** showed a ~ 3-4 µm thick recrystallized layer, with pores primarily located at the bottom of the recrystallized layer, as shown in **Figure 6a**. SEM-EDS mapping (**Figure 6b,c**) showed that the recrystallized region was Cr depleted compared the alloy composition. A STEM cross-section (**Figure 7a**) taken from the brighter region showed similar microstructure with a ~ 10 µm thick sub-surface layer with grains on the order of a few micrometers in size and large pores present through the recrystallized region. EDS mapping (**Figure 7b,c**) showed that the Cr concentration formed a distinct compositional step at the interface between the bicontinuous structure containing nearly no Cr and the underlying alloy containing 30 % Cr.

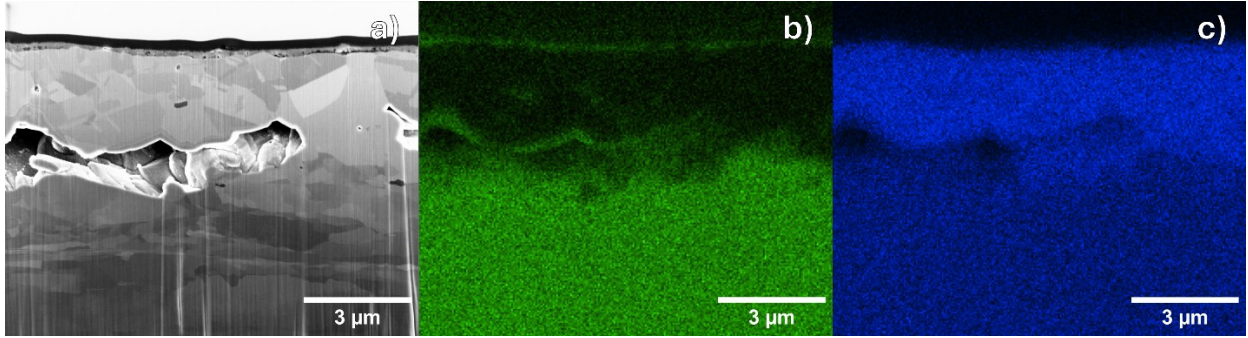

*Figure 6.* FIB cross-section from the dark region of the corrosion-tested sanded surface of the Ni-30Cr alloy, with corresponding EDS elemental maps for (b) Cr and (c) Ni.

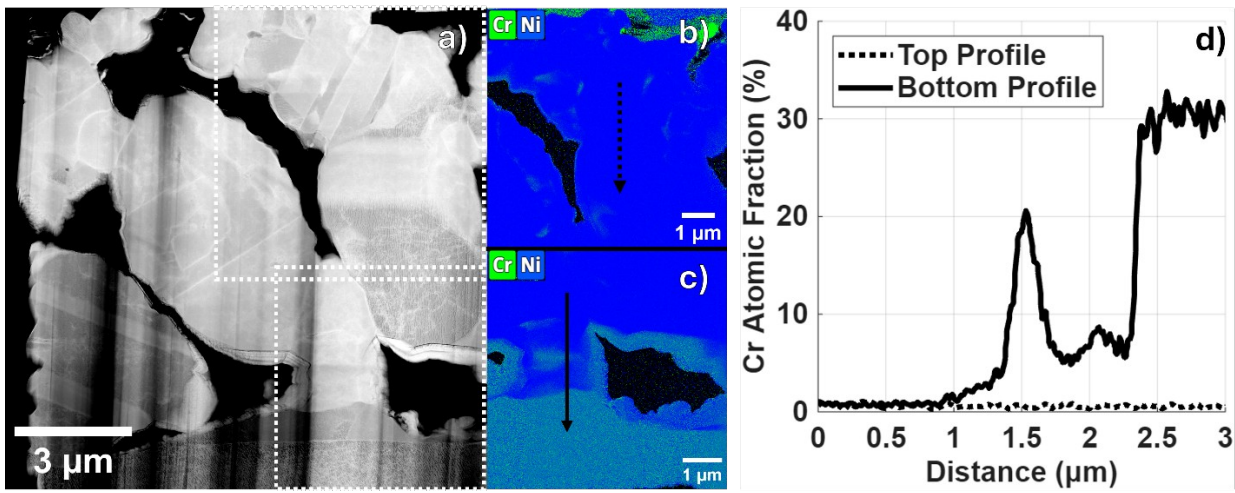

*Figure 7.* (a) HAADF-STEM image of a cross-section from the corrosion tested sanded surface of Ni-30Cr. (b) Corresponding EDS elemental map of the top region highlighted in (a). (c) EDS elemental map of the bottom region highlighted in (a). (d) Cr composition profiles along the red arrows shown in (b, c) expressed as atomic fractions.

## Discussion

The two surface finishes of the Ni-30Cr alloy, EP and sanded, exhibited drastically different post-corrosion morphologies and chemistries. The following discussion compares the present observations to prior reports on selective Cr dissolution in Ni-Cr systems under molten salt exposure [13,15,18,23,32,42], and discusses the role of migrating alloy grain boundaries on the observed corrosion responses.

The corroded EP Ni-30Cr surface presented two distinct behaviors: at the grain boundary/surface intersection and away from grain boundaries. Away from grain boundaries, no solute loss and no concentration gradient was observed. We note here that lattice diffusion of Cr is limited. Using literature lattice diffusion parameters for Ni-Cr alloys ($D_0$ = 4.8 cm$^2$s$^{-1}$, $E_a$ = 291 kJ mol$^{-1}$) [43], the predicted Cr diffusion distance for a 96 hours long exposure is only ~2 nm. The observations suggest that both Cr and Ni dissolved via a layer by layer mechanism, as shown by the crystallographic facets observed on the surfaces of grains of different orientations (**Figure 4c,d**). While Cr is expected to dissolve more easily than Ni because of a higher Gibbs free energy of formation of the metal chloride ($NiCl_2$) [46,47]. the observed behavior suggests that Ni also dissolves but at a slower rate and its dissolution is the rate limiting step for the observed layer by layer dissolution. We note that similar faceted regions and surface morphologies were reported for a Ni-20Cr alloy tested in FliNaK salt at 600 °C under potentiostatic polarization [44].

At the grain boundary/surface intersections, the observed dissolution behavior was consistent with DIGM, that is the coupling between grain boundary migration and solute grain boundary diffusion. Here, DIGM enabled Cr transport to the alloy surface and complete depletion, leaving behind nearly pure Ni grains. The topology of the pure Ni region in **Figure 5**, while rather unexpected, may be explained as follows. The nearly vertical alloy grain boundary migrated via DIGM contributing to the formation of pure Ni regions. The left to right migration was indicated by the position shift at the interface between the alloy and the pure Ni region. However, the sole migration of the vertical grain boundary cannot explain the formation of such large pure Ni grain to the left of the grain boundary - a region that extends beyond the migrated zone - or the formation of a pure Ni grain to the right of it – a region not yet swept by the moving grain boundary. Instead, the concomitant DIGM of the nearly horizontal LAGBs is needed to explain the formation of the Ni grains. The LAGBs migrated downward as Cr diffused laterally along them to the surface in contact with the salt. The origin of the LAGBs is not clear. They may be pre-existing; they may also be the result of injected vacancies during Cr dissolution and/or dislocations left behind by the vertical grain boundary in order to accommodate the volume change associated with the chemical change from Ni-30Cr to pure Ni [29]. The proposed evolution of the EP surface is presented schematically in **Figure 8**.

The height difference between the pure Ni regions above the alloy grain boundaries and the grain interiors (**Fig. 5a**) confirms that more material dissolved above the grain interiors than above the grain boundaries. The amount of dissolved Cr measured by ICP-MS analysis (**Figure 1**) supports the present interpretation (**Figure 5b**) that most of the dissolved Cr comes from the grain interiors. Indeed, the average measured concentration is 1164 ± 187 µg g$^{-1}$ (w.ppm) of Cr recovered from the EP samples, which normalized by the droplet's estimated surface area of ~27 mm$^2$, with an alloy density of 8.36 g cm$^{-3}$, yields ~0.059 mm$^3$ of removed material, corresponding to a depth of ~2.7 µm. This depth is consistent with the height difference between the corroded surface in the grain interiors and the top of the very slow dissolving pure Ni grains (**Figure 5a**). However, the smooth somewhat rounded shape of the pure Ni grains suggests that they were also dissolving, albeit at a much slower rate than the Ni-30Cr grain interiors. Ronne et al. also reported Ni dissolution for a Ni-20Cr corroded at 800 ºC [42].

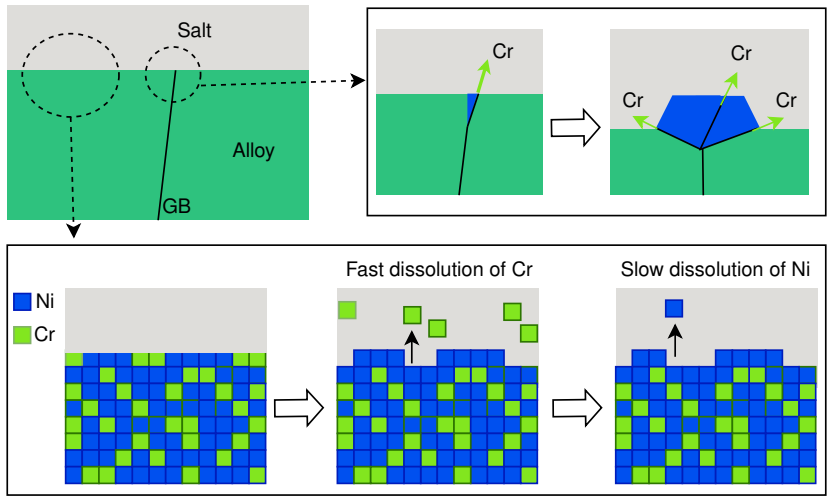

*Figure 8: Schematic of the proposed mechanism for electropolished surface*

Unlike the EP surface, the post-corrosion sanded Ni-30Cr samples displayed a percolating bi-continuous network beneath the sample surface. Comparable morphologies have been reported for Ni-20Cr alloys exposed to chloride based molten salts, where interconnected pores and ligaments formed through selective Cr dissolution [13,15,18,32,23]. Across these studies, the dealloyed ligaments were typically Ni-rich but still contained measurable Cr, with reported values ranging from about 4 at% in ligament cores to about 7–13 wt% at Ni-rich surfaces, and only very thin layers in some cases approaching complete Cr depletion [13,15,18,32,23]. The exception is Ref. [44] where an extensive subsurface layer of pure Ni was observed. The present results also showed complete solute (Cr) depletion throughout the bicontinuous regions. The extent of the measured Cr depletion cannot be explained by lattice diffusion alone. As already mentioned, the predicted Cr diffusion distance for a 96 hours long exposure is only ~2 nm, far below the several micrometer depletion distances observed herein. The estimated grain boundary diffusion, using $D_0 = 65.8$ cm$^2$s$^{-1}$, $E_a = 203$ kJ mol$^{-1}$ [43], predicts a Cr diffusion distance of ~6 µm. While this would be enough to explain depletion along grain boundaries, this alone does not explain depletion away from grain boundaries.

Here, we attribute the Cr dealloying observed on the sanded surface to DIGM. The microstructure observed for the sanded sample can therefore be understood to develop as follows. During the initial moments of exposure, some Cr diffusion might have taken place through the lattice and possibly via pipe diffusion through the network of dislocations created by deformation toward the alloy / salt interface. Because the dissolution rate exceeded the lattice diffusion rate, the surface quickly became depleted of Cr. Concurrently, due to the initial deformation imparted by surface grinding, recrystallization beneath the surface proceeded, producing small grains and a high density of grain boundaries. Recrystallization was confirmed and illustrated in **Figure 9** for a separate sample annealed in an Ar atmosphere for 96 hours. The exact timescale for recrystallization in the corroded regions remains to be clarified. However, Dudova et al. [45] showed that recrystallization of a Ni-20Cr alloy can take place within 1 hour at 500 °C depending on the level of deformation. It is therefore very plausible that recrystallization took place during the first few hours of corrosion testing at 500 °C in the present case. The newly formed grains rapidly grew with moving grain boundaries being subjected to DIGM and therefore leaving behind Cr depleted regions. Grain boundaries swept through the initially deformed layer and continuously drew Cr from the swept areas to the alloy/salt interface (at the sample surface or pores), where Cr dissolved in the molten salt. It is important to note that grain boundaries are not just acting as fast diffusion paths and that DIGM is needed to explain the observed loss of Cr. The proposed mechanism is presented schematically in **Figure 10**. This process bypasses the restrictions of lattice diffusion because solutes are transported more efficiently along moving boundaries rather than through bulk diffusion and/or along static grain boundaries. The abrupt Cr compositional step observed at the same depth where porosity stops (**Figures 5, 6**) supports the interpretation that DIGM was concomitant to pore formation and, consequently, to Cr dissolution. The estimated volume of the pores, from area fraction of the cross-section images was of the order of 0.3, consistent with the total loss of Cr from the starting Ni-30Cr composition. Similarly to the EP samples where Ni dissolved, some dissolution of pure Ni may also have taken place in the case of the sanded sample (as shown by the ICP-MS data), on the sample surface and on the pore surfaces.

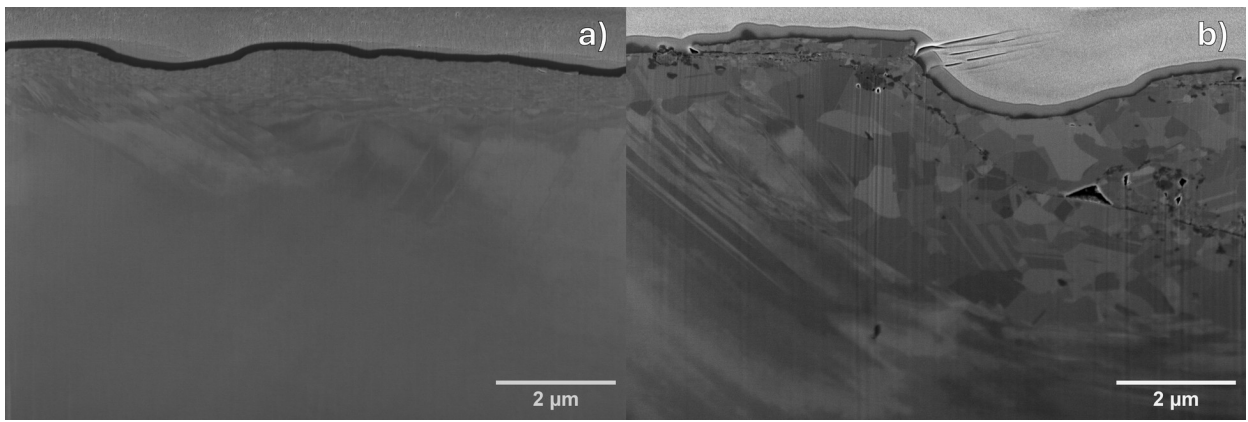

*Figure 9: Cross-section SEM images (secondary electron) of (a) the sanded surface before corrosion exposure, and (b) after thermal annealing at 500 °C for 96 hours (without corrosion exposure).*

Because the present results demonstrate that surface deformation triggers DIGM-assisted dealloying and results in a bi-continuous structure, previously reported corrosion morphologies may be reinterpreted through this lens. Indeed, a number of prior studies used heavily deformed starting microstructures. Ronne et al. [42] reported bicontinuous dealloying in as-draw Ni-20Cr microwires in KCl-MgCl$_2$ at 800 ºC. Liu et al. [15] likewise examined cold-drawn Ni-20Cr microwires corroded in KCl-MgCl$_2$ at 600 ºC and reported percolating pores within a few micrometers sized grains. Yu et al. [13] studied as-rolled Ni-20Cr foils in KCl–MgCl$_2$ and Kcl–NaCl at 700 °C and observed micron-scale porous layers. In these heavily deformed alloys, recrystallization was expected to take place, and we suggest that DIGM may have played a significant role in the observed corrosion responses. Wang et al. [27] reported that grain refinement led to significant acceleration of Cr dissolution in Ni-Cr alloys in (Li,Na,K)F at 700 °C, and while the higher density of grain boundaries can lead to increase Cr loss via the increased density of fast diffusion paths, it is plausible that DIGM could have played a significant role.

The presence of surface deformation can create conditions where solute transport and boundary motion become coupled. The role of DIGM, as highlighted herein, does not preclude other corrosion mechanisms from operating, such as intergranular corrosion and a "percolation dissolution" type mechanism, as studies of less deformed alloys have shown, e.g., [18,23]. Rather, DIGM mediated by deformation and recrystallization, as shown herein, can be considered a complementary mechanism. We also note that the role of DIGM presented here goes beyond the role modeled by Bieberdorf et al. [28] where DIGM only accounts for local Cr depletion around grain boundaries subject to preferential attack.

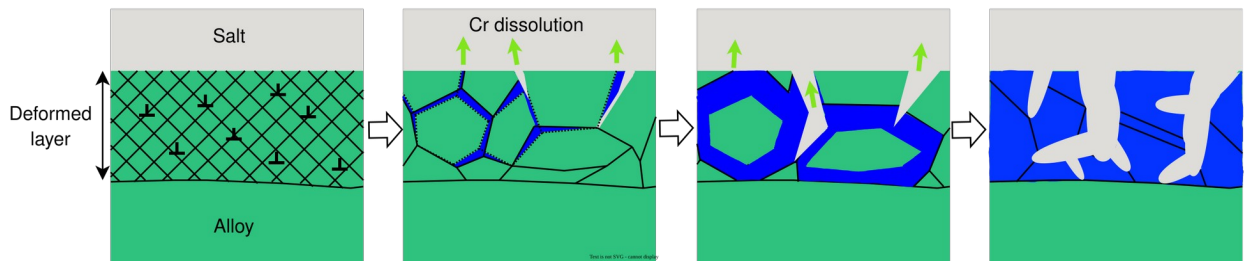

*Figure 10: Schematic of the proposed mechanism for sanded surface*

## Conclusions

We characterized and compared the microstructures of a Ni-30 at.% Cr alloy with two surface finishes, after corrosion in a LiCl-KCl mixture with 2 wt% EuCl$_3$ at 500 °C for 96 hours. Taken together, these observations demonstrate that microstructure via surface preparation can significantly impact the corrosion response. Sanded surfaces, which underwent rapid recovery and recrystallization, exhibited complete selective Cr dissolution and pore formation, while electropolished surfaces showed no selective dissolution, except where alloy grain boundaries intercepted the surface where the formation of small pure Ni grains was noted. These behaviors were rationalized via the contribution of diffusion-induced grain boundary migration, enabling Cr diffusion over large distances. These results underscore the decisive role of surface conditioning on corrosion resistance and degradation pathways in Ni-Cr alloys under molten salt exposure. The findings also suggest that optimizing surface finish, such as by promoting the formation of a Ni enrichment layer, could enhance material lifetime and reliability in molten salt environments. Future investigations should further clarify the interplay between surface preparation, grain boundary dynamics, and molten salt chemistry in order to guide material selection and processing methods for high-performance corrosion resistance.

## Acknowledgments

This work was funded by the National Science Foundation, Division of Materials Research, under Award #2236887. KW acknowledges support from the Department of Energy, Office of Nuclear Energy University Nuclear Leadership Program Graduate Fellowship. The authors thank the staff at the Michigan Center for Materials Characterization for technical support. Authors acknowledge William Minnette for his help during molten salt corrosion studies. The corrosion testing at UW-Madison was supported as part of the Molten Salts in Extreme Environments (MSEE) Energy Frontier Research Center (EFRC), funded by the U.S. Department of Energy, Office of Science, Basic Energy Sciences under contract #DE-SC0012704 to Brookhaven National Laboratory (BNL).

## Data Availability

Data available upon request.